\documentclass[preprint2,times,tighten]{aastex6}
\usepackage{amsmath,amssymb,amstext}
\usepackage[figure,figure*]{hypcap}
\usepackage{newtxmath} %use times font for math

\graphicspath{{./}{figs/}}

\begin{document}

\title{An intrinsic link between  long-term UV/optical variations and X-ray loudness in quasars}
\author{Wen-yong Kang\altaffilmark{1,2}, Jun-Xian Wang\altaffilmark{1,2}, Zhen-Yi Cai\altaffilmark{1,2}, Heng-Xiao Guo\altaffilmark{3}, Fei-Fan Zhu\altaffilmark{1,2}, Xin-Wu Cao\altaffilmark{4,6,7}, Wei-Min Gu\altaffilmark{5,7}, Feng Yuan\altaffilmark{4,7}}
\altaffiltext{1}{CAS Key Laboratory for Researches in Galaxies and Cosmology, University of Science and Technology of China, Chinese Academy of Sciences, Hefei, Anhui 230026, China; kwy0719@mail.ustc.edu.cn, jxw@ustc.edu.cn, zcai@ustc.edu.cn}
\altaffiltext{2}{School of Astronomy and Space Science, University of Science and Technology of China, Hefei 230026, China}
\altaffiltext{3}{Department of Astronomy, University of Illinois at Urbana-Champaign, Urbana, IL 61801, USA}
\altaffiltext{4}{Shanghai Astronomical Observatory, Chinese Academy of Sciences, 80 Nandan Road, Shanghai 200030, China}
\altaffiltext{5}{Department of Astronomy, Xiamen University, Xiamen, Fujian 361005, China}
\altaffiltext{6}{Key Laboratory of Radio Astronomy, Chinese Academy of Sciences, Nanjing 210008, China}
\altaffiltext{7}{SHAO-XMU Joint Center for Astrophysics, Xiamen University, Xiamen, Fujian 361005, China}

\begin{abstract}
Observations have shown that UV/optical variation amplitude of quasars depend on several physical parameters including luminosity, Eddington ratio, and likely also black hole mass. 
Identifying new factors which correlate with the variation is essential to probe the underlying physical processes.
Combining $\sim$ ten years long quasar light curves from SDSS stripe 82 and X-ray data from Stripe 82X, we build a sample of X-ray detected quasars
to investigate the relation between UV/optical variation amplitude ($\sigma_{rms}$) and X-ray loudness.
We find that quasars with more intense X-ray radiation (compared to bolometric luminosity) are more variable in UV/optical.  Such correlation remains
highly significant after excluding the effect of other parameters including luminosity, black hole mass, Eddington ratio, redshift, rest-frame wavelength (i.e., through partial correlation analyses). 
We further find the intrinsic link between X-ray loudness and UV/optical variation is gradually 
more prominent on longer timescales (up to 10 years in the observed frame), but tends to disappear at timescales $<$ 100 days.  This suggests a slow and long-term underlying physical process.  
The X-ray reprocessing paradigm,  in which UV/optical variation is produced by a variable central X-ray emission illuminating the accretion disk, is thus disfavored. 
The discovery points to an interesting scheme
that both the X-ray corona heating and UV/optical variation is quasars are closely associated with magnetic disc turbulence,
and the innermost disc turbulence (where corona heating occurs) correlates with the slow turbulence at larger radii (where UV/optical emission is produced).

\end{abstract}

\section{Introduction}
Active galactic nuclei and quasars, powered by the central accreting supermassive black holes, show 
aperiodical variations from radio waves to X-rays and gamma-rays. 
Investigating the nature of such variations and the underlying physics is one of the main subjects of modern time-domain astronomy.
Particularly, studies on the variations of the UV/optical emission, dominantly produced in the accretion disk, are helpful to probe the underlying physics of the inner accretion process.

Observational studies have established clear (anti-) correlations
between the UV/optical variation amplitude and several known parameters.  UV/optical variability decreases with wavelength (e.g., \citealt{
berk2004ensemble,Wilhite2005,Zuo2012,meusinger2011spectral}), and luminosity, likely driven by the Eddington ratio (e.g., 
\citealt{berk2004ensemble,wilhite2008variability,ai2010dependence,Zuo2012,meusinger2013ultraviolet}). Meanwhile the intrinsic correlation with black hole mass is less clear, after isolating the influence 
of luminosity or Eddington ratio, and rest frame wavelength \citep[e.g.][]{Wold2007, wilhite2008variability, Bauer2009, Zuo2012, meusinger2013ultraviolet, Macleod2010, Kozlowski2016ApJ826}.    
A weak positive correlation between the variation and redshift is also reported, after isolating the effects of rest frame wavelength and luminosity \citep{berk2004ensemble}, but several other studies claimed no significant cosmological evolution \citep[e.g.][]{meusinger2011spectral, Zuo2012, Macleod2010}.

Does the quasar variation correlates with additional observable parameters?   
It is interesting to note that at fixed physical parameters aforementioned, quasars exhibit too large scatter in their variations to be attributed to the sparse light curve sampling and photometric uncertainties
\citep{Macleod2012, Guo2017}, suggesting the variation correlates with additional unknown factor(s). 

The characteristic timescales of quasar UV/optical variations were found consistent with disk thermal timescale, and the variations can thus be attributed to 
thermal fluctuations in the disk likely driven by a turbulent magnetic field \citep[e.g.][]{Kelly2009}. In this scheme the variation amplitude is controlled by the strength of the turbulence.
Identifying additional parameters which correlate with variation is thus essential to study the causes or effects  of the magnetic turbulence. 

Meanwhile, the central compact hot X-ray corona in AGNs is widely believed to be heated by magnetic reconnection in the innermost regions which is directly associated with magnetic turbulence \citep[e.g.][]{Galeev1979, DiMatteo1998}.
Searching for direct observational evidences for this scheme however is rather challenging.  
We speculate there exists an observable link between X-ray radiation and UV/optical variations, i.e., in quasars with stronger accretion disk turbulence (stronger UV/optical variation),
the corona heating is more efficient (higher X-ray power).

SDSS Stripe 82, a 290 deg$^2$ equatorial field of sky, 
has been repeatedly scanned $\sim$ 60 times in the $ugriz$ bands by the Sloan Digital Sky Survey (\citealt{Sesar2007}). 
\cite{Macleod2012} presented recalibrated $\sim$10 yr long SDSS $ugriz$ light curves for 9275 spectroscopically confirmed quasars in Stripe 82,
for most of which measurements of black hole mass, absolutely magnitude (K-corrected) and bolometric luminosity are available from \cite{shen2011catalog}.
Stripe 82X, an X-ray survey with Chandra and XMM-Newton observations, covers 31.3 deg$^2$  overlapping the Stripe 82 field (\citealt{Lamassa2016, ananna2017agn}).
A catalog of 6181 unique X-ray sources has been released \citep{Lamassa2016}, enabling us for the first time to explore the relation between X-ray emission and UV/optical variations in a large quasar sample. 

In \S2 we present the cross-matched quasar sample, along with the measurements of variation amplitude for these quasars. 
We perform correlation analyses in \S3 to reveal the intrinsic correlation between UV/optical variation and X-ray loudness, and a positive and statistical robust correlation is reported.
Discussion on this new discovery is given in \S4. 
Throughout this work, cosmological parameters of $H_0=70 km\cdot{}s^{-1}\cdot{}Mpc^{-1}$, $\Omega_m=0.3$ and $\Omega_{\Lambda}=0.7$ are adopted. 

\section{Quasar Sample}

\cite{shen2011catalog} presented physical properties (including black hole mass $M$, bolometric luminosity $L_{bol}$, and Eddington ratio $\dot{m}$) of 105,783 quasars in the SDSS DR7 quasar catalog.  We cross-match those quasars with  the optical counterparts of the Stripe 82X X-ray source catalog \citep{ananna2017agn},  using a matching radius of 0.7\arcsec. A total of 679 unique matches are found \footnote{
 There are 15 quasars which are associated with 2 X-ray sources, for which we choose the X-ray counterpart closest to the optical position.}. 
We obtain their $ugri$ band light curves from \cite{Macleod2012}. 
We drop $z$ band light curves in which the photometric uncertainties are significantly larger and the intrinsic variations of quasars are considerably weaker comparing with the other 4 bands. 
Light curve data points with photometric uncertainties $>$ 0.2 mag ($\sim$ 0.9\% of all data points), mostly due to poor observing conditions,  are also dropped. 
To ensure accurate measurement of the variation amplitudes with the sparsely sampled light curves, we only include quasars with at least 20 photometric data points in each of the light curves. 
We further exclude quasars with redshift $>$ 1.9 from this study, for which the black hole mass derived from CIV line could be significantly biased \citep[e.g.][]{Coatman2016, Coatman2017}.
The final sample contains 499 quasars, with soft (0.5 -- 2 keV), hard (2.0 -- 10.0 keV), and total (0.5 -- 10.0 keV) band X-ray fluxes available for 492, 360, and 499 of them, respectively.

  For each quasar, the intrinsic variation amplitude in each band is measured with the excess variance $\sigma_{rms}$ (\citealt{2003MNRAS.345.1271V}, see also \citealt{Zuo2012}, \citealt{Sesar2007})
  \begin{align}
  \sigma_{rms}^2=\frac{1}{N-1}\sum(X_i-\bar{X})^2 - \frac{1}{N}\sum\sigma_i^2
   \label{e1}
  \end{align}
 where $N$ is the number of photometric measurements, $X_i$ the observed magnitude, $\bar{X}$ the average magnitude, and $\sigma_i$ the photometric uncertainty of each observation. 
 In case of no intrinsic variation, the expected value of  $\sigma_{rms}^2$ is zero, with a statistical uncertainty of 
  \begin{align}
  err(\sigma_{rms}^2)=\sqrt{\frac{2}{N}}\times\frac{1}{N}\sum\sigma_i^2
 \label{e2}
  \end{align}
 due to photometric errors (\citealt{2003MNRAS.345.1271V}). 
 In the following analysis, we assign a 2$\sigma$ upper limit to $\sigma_{rms}^2$ for sources with $\sigma_{rms}^2/err(\sigma_{rms}^2)<2$. 

We note that many studies model quasar light curves with the damped random walk process \citep{Kelly2009, Macleod2010, Zu2013, Kozlowski2010}, 
with two model parameters $\tau$ (the characteristic timescale) and SF$_{\infty}$. However, these parameters can be poorly constrained with SDSS Stripe 82 light curves (due to the limited length
and the sparse sampling, e.g. \citealt{Kozlowski2017}). In this work we simply adopt the standard $\sigma_{rms}$, which describes the variation with a single parameter,  to measure the $\sim$ 10 yrs long variation amplitude of each quasar and study its correlation with other parameters.

\section{Correlation Analyses}

\begin{figure*}%[!h]
  \centering
  \includegraphics[width=\linewidth]{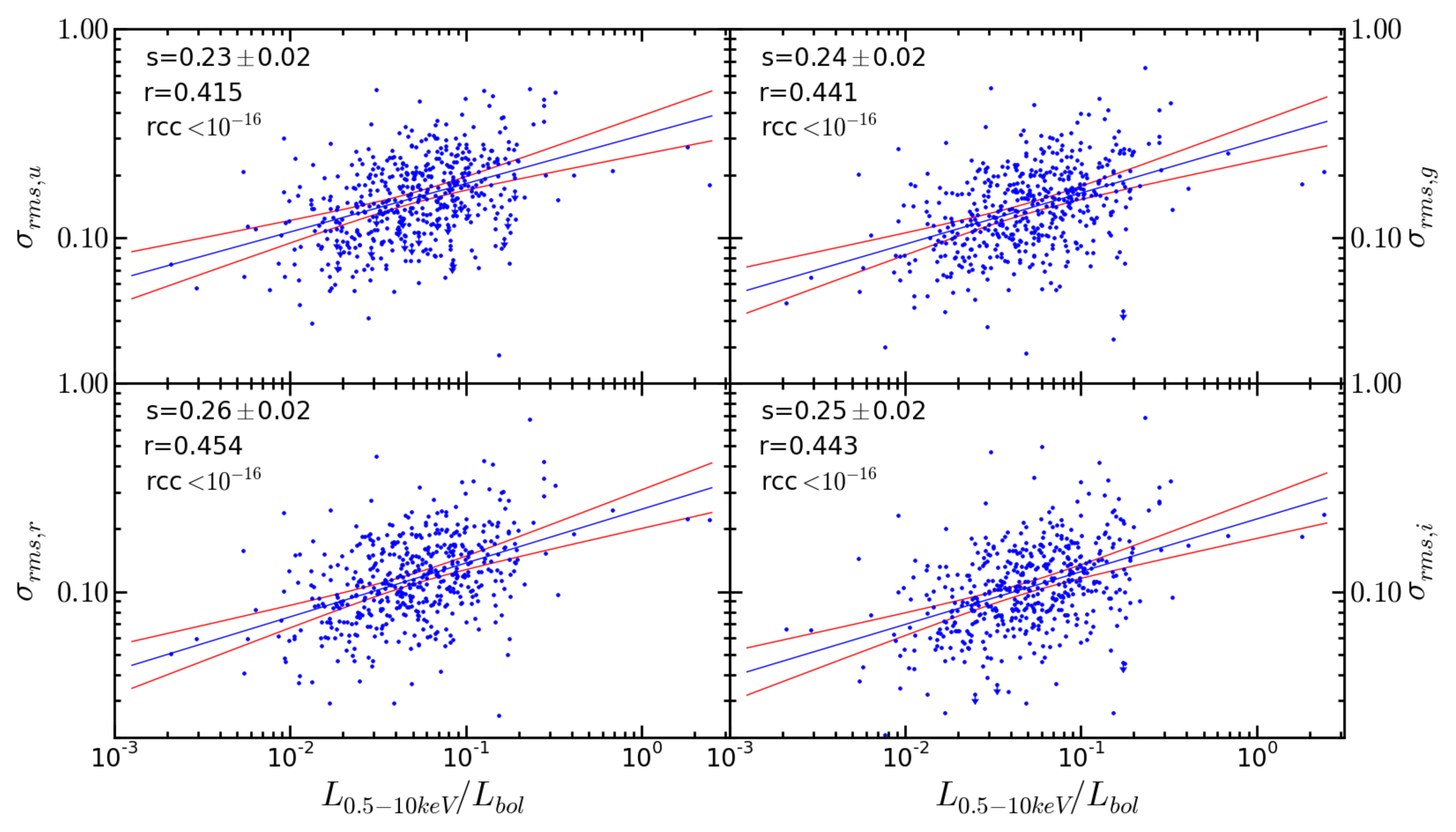}
  \caption{Variation amplitude $\sigma_{rms}$ (in unit of magnitude) in SDSS $ugri$ bands versus X-ray loudness for quasars in SDSS Stripe 82. 
Blue lines plot the best-fit correlations (through simple linear regression) with 3$\sigma$ confidence bands (red lines).
The best-fit linear regression slope $s$, the Spearman's Rank correlation coefficient and the significance level ($r$ and $rcc$) are given in the upper left corner in each panel.
}
  \label{correlation}
\end{figure*}

 We adopt the ratio of X-ray luminosity to bolometric luminosity to represent the relative strength of X-ray emission in each quasar,  which we call X-ray loudness, and investigate its correlation with UV/optical variation amplitudes through simple linear regression: 
 \begin{align}
    \sigma_{rms} \sim (L_{X}/L_{bol})^{s_0}
 \label{e3}
 \end{align}
In Fig. \ref{correlation}  we plot $ugri$ $\sigma_{rms}$ versus $L_{0.5-10 keV}/L_{bol}$, where clear positive correlations are seen. 
Note here we present X-ray loudness based on 0.5 -- 10 keV band X-ray luminosity. 
Using soft and hard X-ray band yield similar results. 
Along with the best-fit linear regression slope $s$, in Fig. \ref{correlation} we also present the Spearman's Rank correlation coefficient and the corresponding significance level ($r$ and $rcc$). 
Note the Spearman's Rank correlation is a non-parametric approach with no prior assumption on either the data distribution or the form of the relationship between two quantities. 
Clearly, both approaches (linear regression and Spearman's Rank) yield clear positive correlations between $\sigma_{rms}$ and X-ray loudness, though with considerable scatter

 ($\sim$ 0.2 dex along the vertical axis, see Fig. \ref{correlation}).

\begin{figure*}%[!h]
  \centering
  \includegraphics[width=\linewidth]{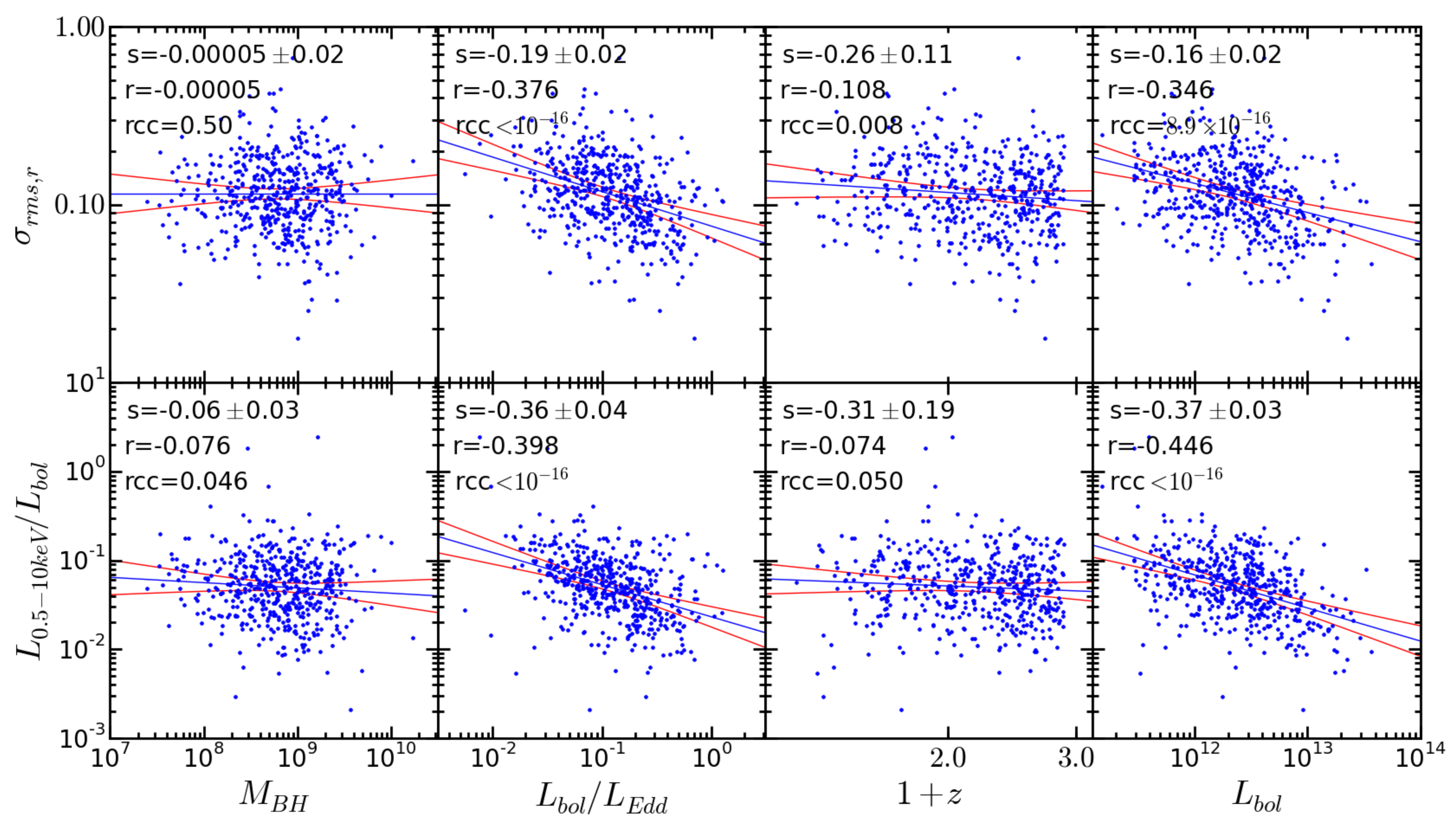}
  \caption{The correlation between $\sigma_{rms}$ (upper panels),  X-ray loudness (lower panels), and black hole mass, Eddington ratio, redshift, and bolometric luminosity. 
  Symbols and lines are the same as in Fig. \ref{correlation}.
  }
  \label{commonrelation}
\end{figure*}

Such correlations demonstrate sources with relatively stronger X-ray emission tend to be more variable in UV/optical.
However, a solid link between them can not yet be established as it is known that both X-ray loudness and UV/optical variation anti-correlate with luminosity (or Eddington ratio)
(\citealt{Bauer2009,Zuo2012,meusinger2013ultraviolet}, \citealt{meusinger2011spectral, Lusso2010, Lusso2012, Fanali2013}),
thus the correlations we plot in Fig. \ref{correlation} might be just secondary effects.
We perform linear regression to examine whether X-ray loudness and UV/optical variation in our sample correlate with common physical parameters including black hole mass M, 
Eddington ratio $\dot{m}$, redshift z, and $L_{bol}$ (Fig. \ref{commonrelation}). 
We do see that in our sample both $\sigma_{rms}$ and X-ray loudness significantly and similarly anti-correlate with $L_{bol}$, and $\dot{m}$, and marginally anti-correlate with redshift.
Meanwhile there is no apparent correlation between $\sigma_{rms}$ and  M, and  a marginal negative correlation between X-ray loudness and M.\footnote{
 However revealing the intrinsic correlation with M
requires isolating the influence of luminosity/Eddington ratio and redshift, which is beyond the scope of this work.
}

Partial correlation analyses were thus required to investigate whether there is intrinsic correlation between UV/optical variation and X-ray loudness,  by controlling the effect of $\dot{m}$, $M$ and redshift. Note that as 
$\dot{m}$ is the ratio of $L_{bol}$ and $M$, the effect of $L_{bol}$ is also simultaneously controlled during the analyses. 
Replacing $\dot{m}$ with $L_{bol}$ during the analyses does not alter the results. 
The correlation coefficient and the significance level ($r$ and $rcc$) are shown in Table \ref{Tab1}, and the intrinsic correlations are weaker than the apparent ones (Fig. \ref{correlation}), but remain statistically significant.  
Such positive intrinsic correlations show that for two quasars with the same bolometric luminosity, black hole mass and redshift, the one with higher X-ray loudness is more variable in UV/optical.

Multiple linear regression analyses were then run to quantify the relations between UV/optical variations and physical parameters including Eddington ratio, black hole mass, redshift, and X-ray loudness 
 \begin{align}
 %\label{e4}
\sigma_{rms} \sim \dot{m}^aM^b(1+z)^c(L_X/L_{bol})^s
 \label{e4}
 \end{align}
and the best-fit parameters are also given in Table \ref{Tab1}, indicating clear intrinsic correlation between $\sigma_{rms}$ and $L_X/L_{bol}$.
Again, replacing $\dot{m}$ with $L_{bol}$ in the equation would not alter the main results presented in this work.

\begin{figure*}%[!h]
  \centering
  \includegraphics[width=\linewidth]{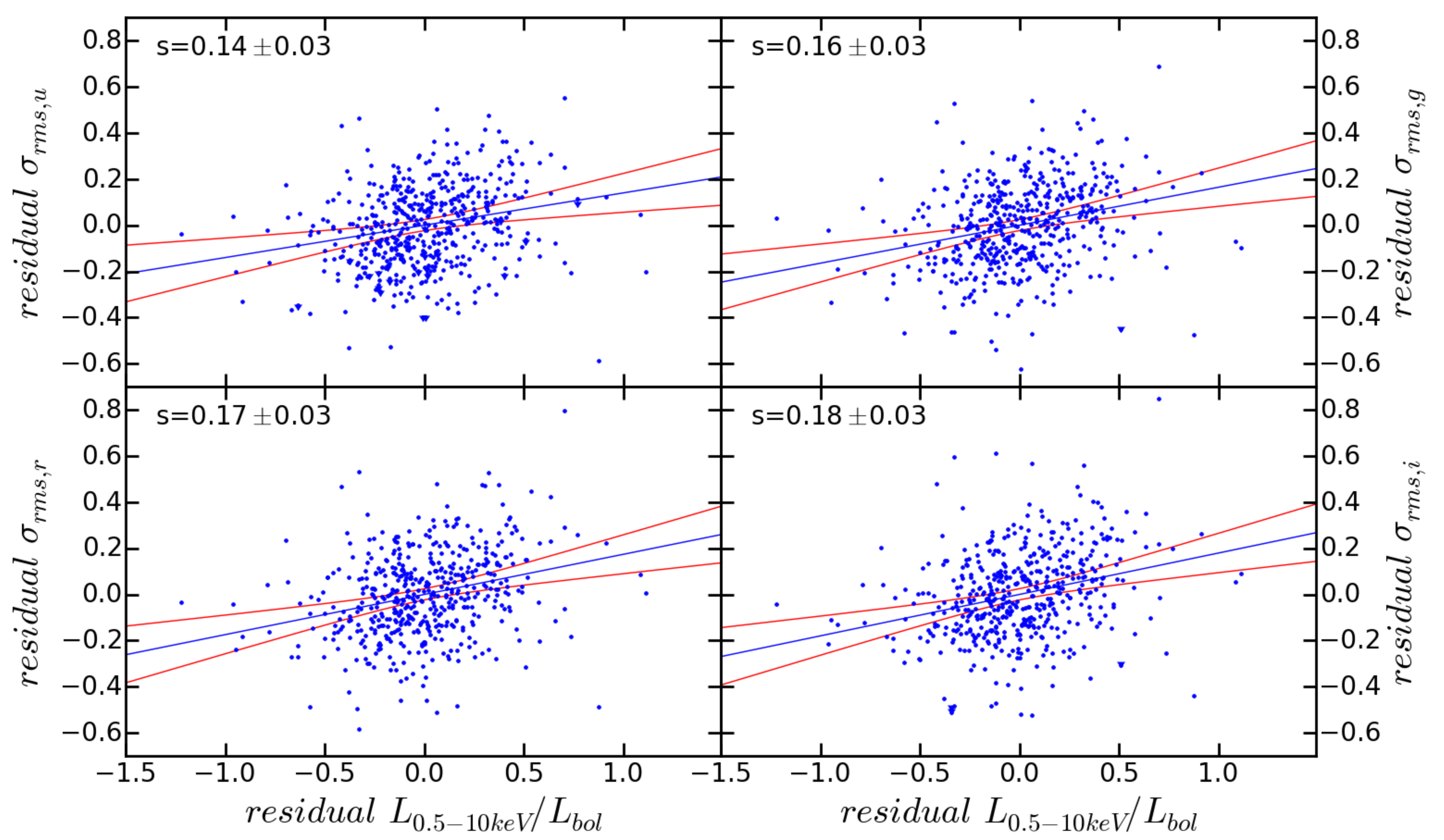}
  \caption{
 The residuals $ugri$ $\sigma_{rms}$  to equation 5 versus the residual X-ray loudness to equation 6. 
 The scatter along the vertical axis is $\sim$ 0.2 dex, similar to that in Fig. \ref{correlation}.
Symbols and lines are the same as in Fig. \ref{correlation}.
}
  \label{residuals}
\end{figure*}

To illustrate the intrinsic correlation between $\sigma_{rms}$ and $L_X/L_{bol}$ obtained in equation \ref{e4}, we 
plot in Fig. \ref{residuals} the correlations between the residuals of equation \ref{e5} and those of equation \ref{e6}. 
\begin{align}
   \sigma_{rms} \sim \dot{m}^{a1}M^{b1}(1+z)^{c1}
    \label{e5}
   \\
  L_X/L_{Edd} \sim \dot{m}^{a2}M^{b2}(1+z)^{c2}
 \label{e6}
  \end{align}
 In Fig. \ref{residuals} the scatter is similar to that in Fig. \ref{correlation}. Such scatter could be due to X-ray flux variation, sparse SDSS photometric sampling in Stripe 82, red noise leakage \citep{Guo2017}, uncertainties in M and L$_{bol}$ measurements, and other unknown parameters which either X-ray loudness or $\sigma_{rms}$ might rely on.

The partial correlation analyses above demonstrate positive intrinsic correlations between single band $\sigma_{rms}$ ($u,g,r,i$) and X-ray loudness.
We note that single SDSS band observes different rest frame wavelength for sources at various redshifts. 
The $\sigma_{rms}$ from 4 bands can be analyzed jointly by assigning a rest-frame wavelength $\lambda_c$ to each measurement of $\sigma_{rms}$ (the central wavelength of the corresponding SDSS band divided by 1+$z$).
We then perform partial correlation analyses between $\sigma_{rms}$ ($\lambda_c$) and X-ray loudness, controlling the effect of 
bolometric luminosity, black hole mass (thus also Eddington ratio), redshift, and rest-frame wavelength. 
The resulted partial correlation coefficients are presented in Table \ref{Tab1}, also showing  statistically significant intrinsic correlation between $\sigma_{rms}$ and X-ray loudness.
 We then run multiple linear regression to quantify the correlations
 \begin{align}
  \sigma_{rms} \sim \dot{m}^a(M_{BH})^b(1+z)^c\lambda_{RF}^d(L_X/L_{bol})^s
 \label{e7}
 \end{align}
 and the derived 
 best-fit slopes are presented in Table \ref{Tab1}.
We however note that caution is needed regarding the significance levels of the partial correlation in the joint analysis,
as four band  $\sigma_{rms}$ measurements of a single quasar are not completely independent to each other.

\section{discussion}

Using SDSS light curves from Stripe 82 and X-ray detections from Stripe 82X, in this work we for the first time examine the correlation between
UV/optical variation amplitudes and X-ray loudness of quasars. Partial correlation analyses demonstrate robust intrinsic correlation between them,
controlling the effects of other fundamental physical parameters including bolometric luminosity, black hole mass,  Eddington ratio, redshift and rest frame wavelength. 
Such intrinsic correlation indicates for quasars with identical $L_{bol}$, $M$, $\dot{m}$ and $z$, the ones with stronger UV/optical variations at identical rest wavelength are X-ray louder, or vice versa.

\subsection{The robustness of the intrinsic correlation between UV/optical variation and X-ray loudness }

We note that our sample is limited to SDSS quasars with Stripe 82X X-ray detections. 
Stripe 82X covers an area of 31.3 deg$^2$, while S82 covers an area of 290 deg$^2$.
Taking the sky coverages into consideration, we estimate an X-ray completeness of $\sim$ 73\% for our sample ( of all $z<$1.9 SDSS quasars with $M$ and $L_{bol}$ measurements and sufficient light curve data points in Stripe 82X)\footnote{In the full 290 deg$^2$ S82 area, there are 6306 $z<$1.9 quasars with $M$ and $L_{bol}$ measurements and at least 20 data points in each light curve. We expect $\sim$ 680 such quasars in the 31.3 deg$^2$ S82X area, and our final sample consists 499 quasars with full X-ray band detection.}.
Would the sample incompleteness in X-ray
produce artificial correlation between $\sigma_{rms}$ and X-ray loudness? 
We perform simulations to address this issue. 
Equation \ref{e5} measures the dependency of $\sigma_{rms}$ (X-ray loudness)
on Eddington ratio, mass and redshift, and similarly Equation \ref{e6} measures the dependency of X-ray loudness against these parameters. 

For each individual quasar, starting from its Eddington ratio, mass and redshift, we calculate its expected 
$\sigma_{rms}$ and X-ray loudness based on the best-fit correlations in Equation \ref{e5} \& \ref{e6}. We then add random Gaussian fluctuations (with variance derived 
from the residuals of Equation \ref{e5} \& \ref{e6}) to the expected $\sigma_{rms}$ and X-ray loudness. Obviously the simulated $\sigma_{rms}$ and X-ray loudness 
show no intrinsic correlation at all (controlling the effects of observed Eddington ratio, mass and redshift), and such fact is confirmed with partial correlation and multiple linear regression analyses.
We then apply an X-ray flux cut to this simulated sample to mimic the effect of an X-ray incomplete sample. The cut was selected to exclude 27\% of the X-ray weak sources in the sample. The resulting incomplete sample does not show any ``intrinsic" correlation between $\sigma_{rms}$ and X-ray loudness.

It's known that the measurements of black hole mass and bolometric luminosity of quasars suffer considerable uncertainties. 
Assuming both $\sigma_{rms}$ and X-ray loudness correlate with luminosity or Eddington ratio but without intrinsic correlation between them,
uncertainties in $M$ and $L_{bol}$ might lead to artificial partial correlation between $\sigma_{rms}$ and X-ray loudness.
We also perform Monte-Carlo simulations to address such effect. 
Again, from Equation \ref{e5} \& \ref{e6} we build artificial samples with no intrinsic correlation between simulated $\sigma_{rms}$ and X-ray loudness. 
We then add random fluctuations to the observed $L_{bol}$ and $M$ for each quasar.  For mass measurement, we adopt a conservative 0.4 dex calibration uncertainty \citep{shen2011catalog}, and add it quadratically to the direct measurement error from \cite{shen2011catalog}. For $L_{bol}$, both a 0.08 dex uncertainty (20\%, to take care of the uncertainty in bolometric correction, \citealt{Richards2006}) and direct measurement error from \citep{shen2011catalog} are included. 
No fluctuation is added to redshift as it has considerably small uncertainty. 
Partial correlation analyses of using the simulated  $\sigma_{rms}$, X-ray loudness, $L_{bol}$ and $M$ however do not yield significant  ``intrinsic"
 correlation between $\sigma_{rms}$ and X-ray loudness.
We conclude that the observed correlation between $\sigma_{rms}$ and X-ray loudness is physical, and can not be attributed to any observational effect.

\subsection{The underlying physics}

Such correlation  seemingly appears consistent with the so-called X-ray reprocessing paradigm (\citealt{1991ApJ...371..541K}), 
in which a variable central X-ray emission illuminates the accretion disk and produces variable reprocessed UV/optical radiation. Though the reprocessing paradigm
can reproduce closely coordinated variations and lags between various bands, it faces severe challenges. 
Energy budget is one of the most prominent challenges as X-ray usually makes up only a small fraction of the bolometric luminosity hence would be insufficient to drive strong enough UV/optical variations 
\citep[e.g.][]{Gaskell2007}.
Furthermore, the UV/optical inter-band lags observed are  $\sim$ 3$\times$  larger than the thin disk theory prediction \citep[e.g.][]{Fausnaugh2016, Jiang2017}. 
The lag between UV and X-ray appears even up to $\sim$ 20 times larger the model prediction \citep{McHardy2017}.
It was also found that UV/optical light curves are inconsistent with X-ray reprocessing in many sources, either showing too smooth variations or no clear correlation with X-ray 
\citep[e.g.][]{Gardner2017, Maoz2002, Gaskell2006}. Most recently, \cite{Zhu2018} pointed out that the reprocessing paradigm is unable to reproduce the observed timescale dependent color variation observed in AGNs \citep{Sun2014,Zhu2016}, and such discrepancy can not be reconciled under the general reprocessing paradigm. 
Considering all these severe challenges to the reprocessing mode, it is unlikely that the UV/optical variations in quasars are caused by X-ray reprocessing.
Thus the intrinsic correlation between UV/optical variation and X-ray loudness in quasars does not necessarily support the reprocessing model. 

\begin{figure}[!h]
  \centering
  \includegraphics[width=\linewidth]{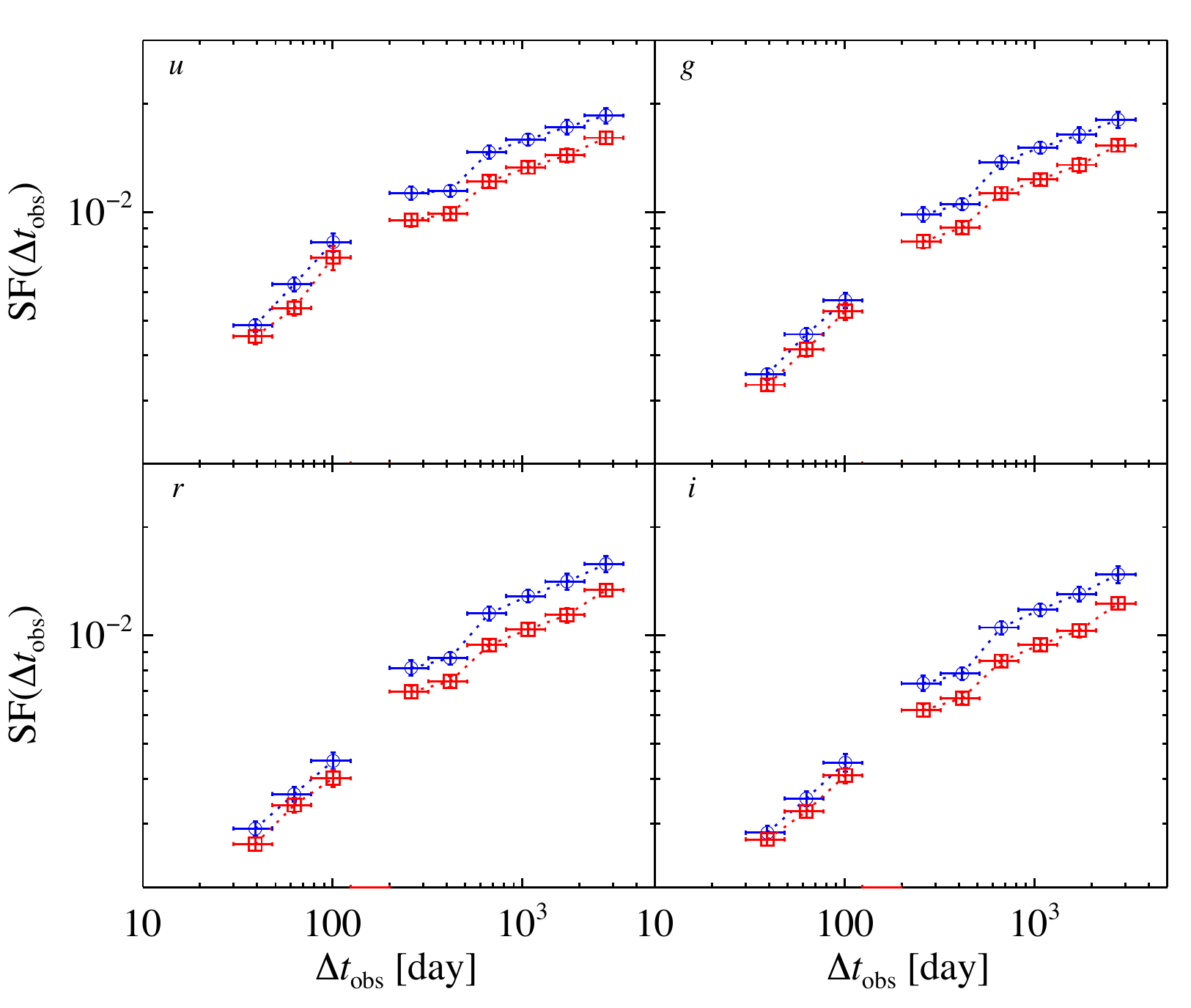}
  \caption{
  Comparison of the $ugri$ ensemble structure function of quasars with relatively higher (blue) and lower (red) X-ray loudness. The errors of the structure function 
  are derived through bootstrapping the samples. The gap around 200 days is due to the lack of timescale coverage of SDSS photometric observations.
  }
  \label{sf}
\end{figure}

As we noted earlier, $\sigma_{rms}$ measures the $\sim$ 10 years long variation amplitudes with a single parameter. To further probe the correlation between
X-ray loudness and UV/optical variation at various timescales, we divide our quasar sample (with full band X-ray detection) equally into two subsamples according to their residual X-ray loudness to equation \ref{e6}, namely X-ray louder and X-ray fainter sample, respectively. 
K-S test confirms the two subsamples have statistically indistinguishable distributions of redshift, black hole mass, bolometric luminosity and Eddington ratio.
Following \citet{diClemente1996} and \citet{Zhu2016}, we derive the ensemble structure function for each sample, i.e., 
\begin{equation}
	\mbox{SF}(\tau|\tau_{\rm min} < \tau < \tau_{\rm max}) = \sqrt{\frac{\pi}{2} <|m_i - m_j|>^2 - <\sigma^2_i + \sigma^2_j>},
\label{e8}
\end{equation}
where, for a given broad-band light curve, $m_i$ and $\sigma_i$ are the observed magnitude and error at $t_i$ epoch, respectively, $<...>$ denotes averaging over all $(m_i, m_j)$ or $(\sigma_i, \sigma_j)$ observational pairs satisfying $\tau_{\rm min} < |t_i - t_j| < \tau_{\rm max}$, and the typical timescale $\tau$ is the logarithmic average between the minimal $\tau_{\rm min}$ and maximal $\tau_{\rm max}$ boundaries. 
Similarly, data points with photometric uncertainties greater than 0.2 mag are excluded. 
As the light curve of each quasar was similarly sampled in the observed frame, we present the ensemble structure functions as a function of timescale in the observed frame. 
Contrarily using rest frame would yield biased SF, e.g., with the data point at longest rest frame timescale bin dominated by low-z sources.

The derived $ugri$ structure functions are plotted in Fig. \ref{sf}, in which we can clearly see stronger variation of the X-ray louder sample, particularly at long timescales, consistent with the detection of
an intrinsic correlation between X-ray loudness and $\sigma_{rms}$. It is interesting to note that the difference in SF between two subsamples is gradually more prominent at longer timescales,
and diminishes (or even disappears) at $\tau$ $<$ 100 days. 
This indicates the physical link between X-ray loudness and UV/optical variation occurs on long timescales. 
This also disfavors the X-ray reprocessing paradigm as in which a fast physical link is involved (the photon travel time from the central X-ray source to
the UV/optical emitting accretion disc is a couple of days to weeks for quasars).

Instead, observations have shown that quasar variations, 
with characteristic timescales consistent with disk orbital or thermal timescales,
 can be attributed to disk thermal fluctuations driven by magnetic turbulence \citep{Kelly2009}.
Inhomogeneous disk models accounting such fluctuations appear well consistent with observations. The original model proposed by \cite{Dexter2011} is able to match the disk size with micro-lensing observations but larger than thin disk theory prediction. More excitingly,  the revised inhomogeneous disk models by \cite{Cai2016, Cai2018} can further explain the observed timescale dependent color variations, and inter-band coordinations/lags without light echoing.  

The intrinsic correlation between UV/optical variation and X-ray loudness discovered in this work thus indicates that 
for quasars with stronger disk turbulence, more energy can be dissipated into the corona to produce X-ray emission.
This can be naturally interpreted under the scheme that the X-ray corona in AGNs is heated through magnetic reconnection which is also associated with magnetic turbulence. 
The turbulence propagation along the disc (either inward or outward or both) would be able to link the inner corona heating to outer disc slow turbulences. 
The propagation takes much longer time comparing with X-ray reprocessing. 
However, the exact mechanism of the propagation is yet unclear.
For instance, the pressure  timescales for quasars could be decades to centuries, 
but the detection of changing look quasars suggests the propagation may be on a timescale of 1 -- 10 years, similar to the timescale considered here.
Note the upper hot layer of the disc (where the sound speed could be much larger) and the ultra fast disc outflow may also play a role in the propagation \citep{Cai2018}.
Nevertheless, such long term process(es) (comparing with reprocessing) could naturally explains the fact that the intrinsic link between X-ray loudness and UV/optical variation is gradually stronger at longer timescales  (Fig. \ref{sf}).
We note that even if the process occurs on longer timescales than observed, it may still yield an intrinsic correlation with a sufficiently large sample, and such effect may also contribute to the
scatter in the correlation between X-ray loudness and UV/optical variation.

The multiple linear regression yields $\sigma_{rms}~\sim~(L_X/L_{bol})^{\sim0.16}$ (controlling the effects of other physical parameters).
Analyzing using $\sigma_{rms}$ as the independent variable and $L_X/L_{bol}$ as a dependent variable yields 
$L_X/L_{bol}~\sim~(\sigma_{rms})^{\sim0.43}$. A bisector slope for the correlation between $\sigma_{rms}$ and $L_X/L_{bol}$ 
is $\sim$ 0.78, suggesting a close-to-linear intrinsic relation between them. 
It's interesting to note that the intrinsic correlation between $\sigma_{rms}$ and X-ray loudness appears slightly stronger at longer wavelength (see Table \ref{Tab1}).
This might be due to variation of dust attenuation along the line of sight to some quasars, which would produce flux variations in additional to the 
intrinsic one, mainly at shorter wavelengths.
We also note that the intrinsic correlation between $\sigma_{rms}$ and hard band X-ray loudness is weaker comparing with soft and full band. This is mainly because the hard X-ray sample is considerably smaller than the soft and full band samples. We build a sample with both soft and hard X-ray detections and find no obvious difference in the intrinsic correlation. We do not find a significant intrinsic correlation between $\sigma_{rms}$ and X-ray hardness ratio either. 

Various studies have detected clear anti-correlation between X-ray loudness and luminosity/Eddington ratio of AGNs, indicating highly accreting sources dissipate relatively less energy in the corona \citep[e.g.][]{Vasudevan2009,Lusso2010,Lusso2012,Fanali2013}. 
It's a fundamental question to understand what process controls the fraction of energy dissipated into the corona.
The discovery in this work reveals that the turbulence is (one of) the driven factor(s) behind these correlations.

\section*{Acknowledgement}

 We thank the anonymous referee for constructive suggestions that help to improve the manuscript. 
This work is supported by National Basic Research Program of China (973 program, grant No. 2015CB857005) and National Science Foundation of China (grants No. 11421303 $\&$ 11503024). J.X.W. thanks support from Chinese Top-notch Young Talents Program, 
and CAS Frontier Science Key Research Program (QYZDJ-SSW-SLH006).
Z.Y.C. acknowledges support from the China Postdoctoral Science Foundation (grant No. 2014M560515) and the Fundamental Research Funds for the Central Universities.

\begin{table*}[!h]
  \centering
  \label{tab1}
  \caption{Partial correlation coefficients and multiple linear regression slopes between $\sigma_{rms}$ and other physical parameters}
  \begin{tabular}{c|c|c|c|c|c|c|c|c}
    \hline
      $\sigma_{rms}$ & r & rcc & $\dot{m}$ (a) & M (b) & 1+z (c) & $\lambda_c$ (d) & $L_X/L_{bol}$ (s)& X-ray band  \\
%       (1)&(2)&(3)&(4)&(5)&(6)&(7)&(8)&(9)\\
    \hline
       u&0.230&$1.4\times{}10^{-7}$&-0.20$\pm$0.03&-0.09$\pm$0.03&0.15$\pm$0.14&&0.13$\pm$0.02\\
    \cline{1-8}
       g&0.276&$2.5\times{}10^{-10}$&-0.18$\pm$0.03&-0.06$\pm$0.03&0.28$\pm$0.14&&0.16$\pm$0.02\\
    \cline{1-8}
       r&0.283&$8.7\times{}10^{-11}$&-0.18$\pm$0.03&-0.07$\pm$0.03&0.11$\pm$0.14&&0.17$\pm$0.02 & soft\\
    \cline{1-8}
       i&0.290&$3.2\times{}10^{-11}$&-0.16$\pm$0.03&-0.05$\pm$0.03&-0.06$\pm$0.14&&0.17$\pm$0.02\\
    \cline{1-8}
       u+g+r+i&0.269&$<10^{-16}$&-0.18$\pm$0.02&-0.07$\pm$0.02&-0.42$\pm$0.08&-0.54$\pm$0.03&0.16$\pm$0.01\\
    \hline
       u&0.180&$3.1\times{}10^{-4}$&-0.19$\pm$0.04&-0.10$\pm$0.04&0.20$\pm$0.17&&0.11$\pm$0.03\\
    \cline{1-8}
       g&0.227&$7.2\times{}10^{-6}$&-0.18$\pm$0.04&-0.07$\pm$0.04&0.28$\pm$0.16&&0.14$\pm$0.03\\
    \cline{1-8}
       r&0.262&$2.6\times{}10^{-7}$&-0.16$\pm$0.04&-0.07$\pm$0.04&0.04$\pm$0.17&&0.17$\pm$0.03 & hard\\
    \cline{1-8}
       i&0.279&$4.2\times{}10^{-8}$&-0.12$\pm$0.04&-0.04$\pm$0.04&-0.12$\pm$0.17&&0.18$\pm$0.03\\
    \cline{1-8}
       u+g+r+i&0.237&$<10^{-16}$&-0.16$\pm$0.02&-0.07$\pm$0.02&-0.44$\pm$0.09&-0.55$\pm$0.04&0.15$\pm$0.02\\
    \hline
       u&0.229&$1.3\times{}10^{-7}$&-0.20$\pm$0.03&-0.09$\pm$0.03&0.16$\pm$0.14&&0.14$\pm$0.03\\
    \cline{1-8}
       g&0.270&$4.6\times{}10^{-10}$&-0.18$\pm$0.03&-0.07$\pm$0.03&0.30$\pm$0.14&&0.16$\pm$0.03\\
    \cline{1-8}
       r&0.280&$1.0\times{}10^{-10}$&-0.18$\pm$0.03&-0.08$\pm$0.03&0.12$\pm$0.14&&0.17$\pm$0.03 & full\\
    \cline{1-8}
       i&0.285&$4.9\times{}10^{-11}$&-0.15$\pm$0.03&-0.06$\pm$0.03&-0.04$\pm$0.14&&0.18$\pm$0.03\\
    \cline{1-8}
       u+g+r+i&0.266&$<10^{-16}$&-0.18$\pm$0.02&-0.07$\pm$0.02&-0.40$\pm$0.08&-0.54$\pm$0.03&0.16$\pm$0.01\\
    \hline
  \end{tabular}  
  \\
  \raggedright
  This table lists the best-fit multiple linear regression slopes of Equation \ref{e4} (for band $u$, $g$, $r$, and $i$), and of Equation \ref{e7} (for $u+g+r+i$).
 Here r and rcc represent Spearman's correlation coefficient and significance level of the intrinsic correlation between $\sigma_{rms}$ and $L_X/L_{bol}$ (controlling the effects of other parameters). 
  \label{Tab1}
\end{table*}

\bibliography{qv_180620.bbl}
%\bibliography{ref}

\end{document}